\documentclass[runningheads]{llncs}
\usepackage[T1]{fontenc}
\usepackage{graphicx}
\usepackage[ruled,linesnumbered]{algorithm2e}
\usepackage{float}
\usepackage{amsmath}
\usepackage{amssymb}

\usepackage{cancel}
\usepackage{placeins}

\usepackage{CJK}
\usepackage{setspace}
\usepackage{multicol} 
\usepackage{multirow}
\usepackage{booktabs}  
\usepackage{threeparttable}
\usepackage{algpseudocode} 
\usepackage{subfigure}
\usepackage{xcolor}
\begin{document}

\title{RREH: Reconstruction Relations Embedded Hashing for Semi-Paired Cross-Modal Retrieval}
\author{Jianzong Wang\inst{1 \dagger}, Haoxiang Shi\inst{1,2 \dagger}, Kaiyi Luo\inst{1}, Xulong Zhang\inst{1}\thanks{Corresponding author: Xulong Zhang (zhangxulong@ieee.org)\\$\dagger$ Equal Contributions}, Ning Cheng\inst{1} and Jing Xiao\inst{1}}
\authorrunning{J. Wang et al.}
\titlerunning{RREH: Reconstruction Relations Embedded Hashing}
%
\institute{Ping An Technology (Shenzhen) Co., Ltd., Shenzhen, China\and
University of Science and Technology of China, Hefei, China}
\maketitle              
\begin{abstract}
Known for efficient computation and easy storage, hashing has been extensively explored in cross-modal retrieval. The majority of current hashing models are predicated on the premise of a direct one-to-one mapping between data points. However, in real practice, data correspondence across modalities may be partially provided. In this research, we introduce an innovative unsupervised hashing technique designed for semi-paired cross-modal retrieval tasks, named Reconstruction Relations Embedded Hashing (RREH). RREH assumes that multi-modal data share a common subspace. For paired data, RREH explores the latent consistent information of heterogeneous modalities by seeking a shared representation. For unpaired data, to effectively capture the latent discriminative features, the high-order relationships between unpaired data and anchors are embedded into the latent subspace, which are computed by efficient linear reconstruction. The anchors are sampled from paired data, which improves the efficiency of hash learning. The RREH trains the underlying features and the binary encodings in a unified framework with high-order reconstruction relations preserved. With the well devised objective function and discrete optimization algorithm, RREH is designed to be scalable, making it suitable for large-scale datasets and facilitating efficient cross-modal retrieval. In the evaluation process, the proposed is tested with partially paired data to establish its superiority over several existing methods.
\end{abstract}

\section{Introduction}
The rapid expansion of multimedia content has ignited curiosity in the field of cross-modal search, which entails locating analogous entries across various forms of data representation. For example, a search engine returns videos with text inputs. However, the data heterogeneity of different  modalities complicates similarity measurement, making efficient and accurate search among large-scale datasets difficult \cite{UAPMH}. To address this problem, various Approximate Nearest Neighbor (ANN)  methods have been proposed, of which hashing is widely studied \cite{24a}. Over the past decades, many efficient shallow and deep hashing methods have been proposed \cite{zy1,24yun}. The operational principle of hashing techniques involves mapping diverse types of data onto a common space defined by Hamming distances. In this way, sample similarities can be measured by simple XOR operation, greatly improving the training and retrieval efficiency.

Models for cross-modal hashing can typically be categorized into two main groups: those that employ supervised techniques \cite{DAH,BATCH} and those that utilize unsupervised methodologies \cite{RUCMH,DAEH,zy2,pddh}. Supervised hashing utilizes labels as semantic information to generate robust binary codes and improve the retrieval performance. Some typical researches include scalable asymmetric discrete cross-modal hashing (BATCH) \cite{BATCH}. In contrast, unsupervised hashing deals with a more challenging scenario where no label is provided, and they usually attempt to exploit the data structures to learn binary codes. In recent times, advancements in the field of deep learning have significantly propelled the progress of research into deep hashing techniques. Deep hashing methods treat the neural network as a hash function due to its high non-linearity and representation capacity. Typically, approaches that are supervised tend to yield superior results when compared to their unsupervised counterparts. Nonetheless, gathering labeled data can be a laborious process, particularly for vast multimedia datasets. Consequently, our study is centered on the exploration of hashing methods that are unsupervised and applicable across different modalities.

Most existing unsupervised cross-modal hashing methods require full data correspondence to generate unified binary codes. In real applications, fully paired data are hard to obtain and it is often the case that only partial correspondence is given. Within this framework, there exists no straightforward method to develop a cohesive set of binary representations that cater to every modality, thereby addressing the disparity among diverse data types. Such a scenario, characterized by an incomplete pairing of modalities, is often termed as semi-paired data, as discussed in the literature. There are several methods dedicated to semi-paired cross-modal hashing (SPCMH), such as SPDH \cite{SPDH}, UAPMH \cite{UAPMH} and DUMCH \cite{DUMCH}. Some methods like SPDH preserve pairwise similarities into hash codes based on a predefined Laplacian matrix. Nevertheless, there are several limitations in these learning paradigms under the SPCMH setting: 1) Feature discrimination of the learned hash codes cannot be guaranteed with only paired data used for feature alignment, especially when paired information is little. 2) Solving the Laplacian matrix requires huge memory and computational costs with time complexity $O(n^2)$, making it unscalable for large-scale datasets. 3) Most existing SPCMH methods focus on keeping pairwise similarities and neglect high-order data similarities. The global relationships between paired data and unpaired data are hardly explored.

In this paper, an innovative approach is introduced that specifically addresses the under-explored domain of semi-paired data within the context of unsupervised problem for cross-modal retrieval, termed Reconstruction Relations Embedded Hashing, i.e., RREH for short. Previous SPCMH methods keep  similarities among samples with a Laplacian matrix, while RREH explores the high-order reconstruction relations between paired data and unpaired data, which are embedded into the latent subspace by efficient linear reconstruction. In this way, the modality-specific latent discriminative features are captured. RREH is an efficient model with low time complexity, which can be effectively optimized. The key contributions presented in this paper are as follows:
\begin{itemize}
    \item We introduce a novel unsupervised method, named RREH, designed to address semi-paired cross-modal retrieval tasks.
    \item RREH incorporates a novel strategy called reconstructed relationship embedding, which relies on randomly selected anchors to maintain data similarity. This ensures the preservation of high-order reconstructed relationships between data, even in the absence of paired data.
    \item We design a discrete optimization algorithm to optimize RREH, which significantly reduces the complexity of the optimization process and generates more discriminative hash codes by simultaneously learning hash functions and hash codes.
    \item The MAP results conducted on two widely used datasets demonstrate that the RREH achieves cutting-edge performance with respect to both precision and computational efficiency.
\end{itemize}

\section{Proposed Method}
\begin{figure*}[t]
	\centering
	\includegraphics[scale=0.8]{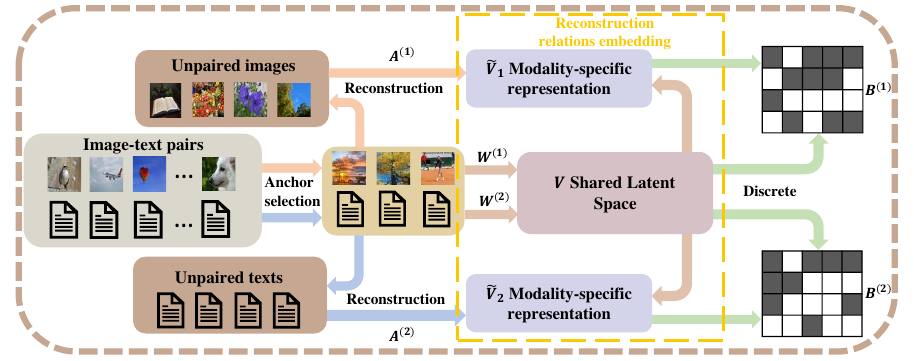}
	\caption{The overall workflow of RREH.}
	\label{workflow}
\end{figure*} 
In this section, the proposed RREH is elaborated, including formulation, optimization and time complexity. The overall workflow of RREH is illustrated in Fig. \ref{workflow}. Firstly, anchors were randomly selected from paired samples for each modality. The linear reconstruction relations can be learned between the anchors and unpaired samples by efficient linear regression, which are then used to reconstruct the modality-specific latent representation, thus preserving the high-order data similarities instead of using pairwise similarities. As demonstrated previously, the cross-modal data similarity is hard to measure and incorporating pairwise similarities is time-consuming. In RREH, we address the two problems with the reconstruction factor, which fully exploits the relationship between paired samples and unpaired ones. The latent representation is then transformed to discriminative binary codes, preserving the high-order data similarities.

\subsection{Notations}
For better elaboration, we list all the necessary notation for model formulation. For the $i^{th}$ modality, the data matrix is defined as $\mathbf{M}^{(i)}\in\mathbb{R}^{d_{i}\times n_i}$, where $d_{i}$ is the dimension and $n_i$ is the data size. Without loss of generality, it is assumed that the first $n_c$ samples in all modalities are paired. To elaborate, let the matrix $\mathbf{M}^{(i)}$ be defined as the concatenation of $\mathbf{M}_{p}^{(i)}$ and $\mathbf{M}_{up}^{(i)}$. Here, $\mathbf{M}_{p}^{(i)}$, which belongs to the space $\mathbb{R}^{d_{i} \times n_c}$, represents the matrix of paired data points. Meanwhile, $\mathbf{M}_{up}^{(i)}$, situated in $\mathbb{R}^{d_{i} \times nu_i}$, signifies the matrix of unpaired data points. The term $nu_i$ corresponds to the quantity of unassociated instances within the $i^{th}$ modality. It is a prevalent assumption that multi-modal datasets are zero-mean, that is, the summation across all data points $\mathbf{m}_{j}^{(i)}$ from 1 to $n_i$ results in the null vector, for $i$ encompassing the range $1, \ldots, m$. Furthermore, the hash code matrix $\mathbf{B}^{(i)}$, specific to the $i^{th}$ modality, consists of binary values and is formatted as an $r \times n_i$ matrix within the set $\{-1, 1\}$. The Frobenius norm of a matrix is indicated by $\|\cdot\|_{F}$, and the operation $\operatorname{Tr}(\cdot)$ is used to extract the trace of a matrix.

\subsection{Reconstruction Factors Learning}
The basic idea of RREH is to bridge paired samples and unpaired samples for preserving similarity within modalities and investigating the correlations across modalities. We assume that unpaired samples of each modality can be linearly reconstructed by paired samples, and the reconstruction factor naturally encodes the relationships among samples. To accelerate the learning process, we randomly select k samples from the paired data as anchor points for all modalities. It should be observed that the composite matrix, denoted as $\mathbf{M}_{p}^{(i)}$, is formulated by the concatenation of two matrices: $\mathbf{M}_{a}^{(i)}$ and $\mathbf{M}_{na}^{(i)}$. Specifically, $\mathbf{M}_{a}^{(i)}$ represents the anchor matrix associated with the $i^{th}$ modality, whereas $\mathbf{M}_{na}^{(i)}$ signifies the matrix comprising the non-anchor, yet paired, data points. It should be noticed that the numbers of anchors for each modality are the same to ease optimization. The reconstruction factors learning can be formulated as:
\begin{equation}
 \min _{\mathbf{R}^{(i)}}\sum_{i=1}^{m}\left(\left\|\mathbf{M}_{na}^{(i)}-\mathbf{M}_{a}^{(i)}\mathbf{R}^{(i)}\right\|_{F}^{2}\right)
\end{equation}
where $\mathbf{R}^{(i)}\in\mathbb{R}^{a\times nu_i}$ is the reconstruction factor for the $i^{th}$ modality.

In an effort to investigate the structure of high-dimensional data, the RREH model employs a kernel technique to enhance the data matrix \cite{SRLCH}. This technique involves a transformation of the original data into a kernel-induced feature space, which is facilitated by the subsequent function:
\begin{equation}
	\phi(\mathbf{m})=\left[\exp \left(\frac{-\left\|\mathbf{m}-\mathbf{m}_{1}\right\|}{2 \delta^{2}}\right), \ldots, \exp \left(\frac{-\left\|\mathbf{m}-\mathbf{m}_{k}\right\|}{2 \delta^{2}}\right)\right]^{T}
\end{equation}
given a set of data points, let $\mathbf{m}_{1}, \mathbf{m}_{2}, \ldots, \mathbf{m}_{k}$ signify a subset of $k$ samples that are chosen at random. Additionally, the term $\delta$ is introduced to represent the bandwidth, which is calculated as the mean Euclidean distance to the $k$ samples from the complement of the training samples within the dataset. So the function to be optimized is:
\begin{equation}
	\min _{\mathbf{R}^{(i)}}\sum_{i=1}^{m}\left(\left\|\phi(\mathbf{M}_{up}^{(i)})-\phi(\mathbf{M}_{a}^{(i)})\mathbf{R}^{(i)}\right\|_{F}^{2}\right)
	\label{first},
\end{equation}
where $\phi(\mathbf{M}_{up}^{(i)})\in\mathbb{R}^{k_i\times nu_i}$ and $\phi(\mathbf{M}_{a}^{(i)})\in\mathbb{R}^{k_i\times a}$ are kernelized data matrices. Apparently, it is a typical least square problem and has a closed-form solution:
\begin{equation}
	\mathbf{R}^{(i)} = \left(\phi(\mathbf{M}_{a}^{(i)})^T\phi(\mathbf{M}_{a}^{(i)}) + \lambda \mathbf{I}\right)^{-1}\phi(\mathbf{M}_{a}^{(i)})^T\phi(\mathbf{M}_{up}^{(i)}).
	\label{A}
\end{equation}
$\lambda$ is the regularization parameter to prevent overfitting. Compared to the Laplacian matrix, preserving similarities with $\mathbf{R}^{(i)}$ requires lower time complexity.
\subsection{Hash Learning}
\subsubsection{Shared Latent Representation Learning.} It is widely theorized that data conveying the same concept should share a consistent latent representation. For the paired samples $\mathbf{M}_{p}^{(i)}$, the shared latent representation can be explored in a fully paired fashion. The following sub-problem is constructed for optimization to generate the shared latent representation:
\begin{equation}
	\min _{\mathbf{W}^{(i)}, \mathbf{V}}
	\sum_{i=1}^{m}\left\|\mathbf{W}^{(i)} \mathbf{\phi(M}_{p}^{(i)})-\mathbf{V}\right\|_{F}^{2},
	\label{second_one}
\end{equation}
where $\mathbf{W}^{(i)}\in\mathbb{R}^{r\times k_i}$ is a hash function. $\mathbf{V}$, residing in $\mathbb{R}^{r \times n_c}$, is designated as the common latent representation across modalities. $\mathbf{V} = [\ddot{\mathbf{V}},\mathbf{\widehat{V}}]$, where $\ddot{\mathbf{V}}$ is the representation for reconstruction anchors, and  $\mathbf{\widehat{V}}$ is the representation for non-anchor paired data. Consequently, the equation referenced as (\ref{second_one}) may be expressed in the following form:
\begin{equation}
	\min _{\mathbf{W}^{(i)}, \ddot{\mathbf{V}},\mathbf{\widehat{V}}}
	\sum_{i=1}^{m}\left\|\mathbf{W}^{(i)} [\phi(\mathbf{M}_{a}^{(i)}),\phi(\mathbf{M}_{na}^{(i)})]-[\ddot{\mathbf{V}},\mathbf{\widehat{V}}]\right\|_{F}^{2}
	\label{second_one_}
\end{equation}

\subsubsection{Construction Relations Embedding.} It is difficult to determine similarities among semi-paired data due to the partiality of paired information. For unsupervised cross-modal hashing, it is crucial to preserve data similarities for better performance. The Laplacian graph is widely adopted for data structure preservation. Nevertheless, constructing the Laplacian matrix with low-level semantic information is inaccurate and time-consuming. RREH suggests embedding high-order reconstruction relations into latent subspace. In this way, the relationships between unpaired data and anchors are kept in the subspace. The relationships between $\mathbf{\widetilde{V}}^{(i)}$ and $\mathbf{V}$ can be bridged with reconstruction factor $\mathbf{R}^{(i)}$. RREH keeps data similarities by assuming that if $\mathbf{M}_{up}^{(i)}$ can be reconstructed by $\mathbf{M}_{a}^{(i)}$, $\mathbf{\widetilde{V}}^{(i)}$ can also be reconstructed by $\ddot{\mathbf{V}}$. To embed the reconstruction relations of high-order data into the latent subspace, we solve the following problem:
\begin{equation}
	\min _{\mathbf{W}^{(i)},\ddot{\mathbf{V}}}
	\sum_{i=1}^{m}\left\|\mathbf{W}^{(i)} \mathbf{\phi(M}_{up}^{(i)})-\ddot{\mathbf{V}}\mathbf{R}^{(i)}\right\|_{F}^{2}.
	\label{second_two}
\end{equation}
RREH avoids computing pairwise similarities and embeds the high-order relationships into the latent representation. From Eq. (\ref{second_two}), it can be seen that the shared latent subspace guides the learning of modality-specific latent representation.

\subsubsection{Hash Code Learning.} Latent representation $\mathbf{V}$ contains  discriminative features of multi-modal data, and It can be perceived as a progressive refinement towards the binary hash codes denoted by the matrix $\mathbf{B}$. To derive $\mathbf{B}$, we address the subsequent optimization challenge:
\begin{equation}
	\label{second_three}
	\begin{aligned}
		&\min _{\mathbf{V},\mathbf{R}^{(i)},\mathbf{\widetilde{B}}_{i},\mathbf{\overline{B}}}
		\sum_{i=1}^{m}\left\|\mathbf{\widetilde{B}}_{i}-\ddot{\mathbf{V}}\mathbf{R}^{(i)}\right\|_{F}^{2} +\left\|\mathbf{\overline{B}}-\mathbf{V}\right\|_{F}^{2}\\
		&s.t.\quad \mathbf{B}^{(i)}\in\{-1,1\}^{r\times n_i},
	\end{aligned}
\end{equation}
where $\mathbf{B}^{(i)} = [\mathbf{\overline{B}},\mathbf{\widetilde{B}}_{i}]$, $r$ is the code length. $\mathbf{\overline{B}} = [\ddot{\mathbf{B}},\mathbf{\widehat{B}}] $ is the shared hash codes, $\mathbf{\widetilde{B}}_{i}$ is the modality-specific hash codes corresponding to unpaired data. This term is introduced to make the latent representation approximate hash codes as much as possible.

\subsubsection{Overall Objective Function.} The overall objective function is expressed as:
\begin{equation}
	\label{overall}
	\begin{aligned}
		\min _{\mathbf{W}^{(i)},\mathbf{V},\mathbf{B}^{(i)}}&\sum_{i=1}^{m}\left(\left\|\mathbf{W}^{(i)} \phi(\mathbf{M}_{p}^{(i)})-\mathbf{V}\right\|_{F}^{2}+\beta\left\|\mathbf{W}^{(i)} \phi(\mathbf{M}_{up}^{(i)})-\ddot{\mathbf{V}}\mathbf{R}^{(i)}\right\|_{F}^{2}\right)\\		
		&+\theta\sum_{i=1}^{m}\left\|[\mathbf{\overline{B}},\mathbf{\widetilde{B}}_{i}]-[\mathbf{V}, \ddot{\mathbf{V}}\mathbf{R}^{(i)}]\right\|_{F}^{2}
		\\
		&s.t.\quad \mathbf{B}^{(i)}\in\{-1,1\}^{r\times n_i},
	\end{aligned}
\end{equation}
where $\beta$ and $\theta$ are tunable hyper-parameters. By optimizing Eq. (\ref{overall}), reconstruction relations are embedded into hash codes, thus preserving semantic information and obtaining robust modality-specific hash functions.

\subsection{Optimization}
Eq. (\ref{overall}) is nonconnex w.r.t all variables due to the binary constraints, thus directly minimizing Eq. (\ref{overall}) is intractable. Some methods \cite{SRSH} adopt the relaxation strategy on the binary codes, which causes huge quantization error. To address this problem, hash codes and hash functions are optimized with the efficient alternate optimization method. Concretely, when a variable is being updated, the remaining variables are fixed.\par
\textbf{Update $\mathbf{W}^{(i)}$.} The following sub-problem is written by eliminating irrelevant variables:
\begin{equation}
	\begin{aligned}
		\min _{\mathbf{W}^{(i)}}\left\|\mathbf{W}^{(i)} \mathbf{\phi(M}_{p}^{(i)})-\mathbf{V}\right\|_{F}^{2}
		+ \beta\left\|\mathbf{W}^{(i)} \mathbf{\phi(M}_{up}^{(i)})-\ddot{\mathbf{V}}\mathbf{R}^{(i)}\right\|_{F}^{2}	
		\label{sub1}.
	\end{aligned}
\end{equation}
The subproblem is convex w.r.t $\mathbf{W}^{(i)}$. Set the derivative of Eq. (\ref{sub1}) to zero, the closed-form solution of $\mathbf{W}^{(i)}$ can be obtained by:
\begin{equation}
	\begin{aligned}
		\mathbf{W}^{(i)} =& \left(\mathbf{V}\phi(\mathbf{M}_{p}^{(i)})^T + \beta\ddot{\mathbf{V}}\mathbf{R}^{(i)}\phi(\mathbf{M}_{up}^{(i)})^T\right)\\
		&\mathbf{\cdot}\left(\phi(\mathbf{M}_{p}^{(i)})\phi(\mathbf{M}_{p}^{(i)})^T+\beta\phi(\mathbf{M}_{up}^{(i)})\phi(\mathbf{M}_{up}^{(i)})^T+\gamma\mathbf{I}\right)^{-1},
	\end{aligned}
	\label{W}
\end{equation}
where $\left(\phi(\mathbf{M}_{p}^{(i)})\phi(\mathbf{M}_{p}^{(i)})^T+\beta\phi(\mathbf{M}_{up}^{(i)})\phi(\mathbf{M}_{up}^{(i)})^T+\gamma\mathbf{I}\right)^{-1}$ is a constant and can be computed before the iteration, and $\gamma$ is a small integer to prevent overfitting.

\textbf{Update $\mathbf{V}$.} It should be noticed that $\mathbf{V}$ is comprised of $\ddot{\mathbf{V}}$ and $\mathbf{\widehat{V}}$, which are optimized separately. By setting the $\ddot{\mathbf{V}}$ to zero, the optimal solution for $\ddot{\mathbf{V}}$ is obtained by
\begin{equation}
	\ddot{\mathbf{V}} = \mathbf{T}\mathbf{Q}^{-1},
	\label{V1}
\end{equation}
where $\mathbf{T} = \sum_{i=1}^{m}(\mathbf{W}^{(i)}\phi(\mathbf{M}_{a}^{(i)})+\beta\mathbf{W}^{(i)}\phi(\mathbf{M}_{up}^{(i)})\mathbf{R}^{(i)T}+\theta\mathbf{\widetilde{B}}_{i}\mathbf{R}^{(i)T}) + \theta\ddot{\mathbf{B}}$, and $\mathbf{Q} = (\beta+\theta)\sum_{i=1}^{m}\mathbf{R}^{(i)}\mathbf{R}^{(i)T}+(m+\theta)\mathbf{I}$. Similarly, the most favorable outcome for the matrix $\mathbf{\widehat{V}}$ is determined as follows:
\begin{equation}
	\mathbf{\widehat{V}} = (\sum_{i=1}^{m}\mathbf{W}^{(i)}\phi(\mathbf{M}_{na}^{(i)})+\mathbf{\widehat{B}})/(m + \theta)
	\label{V2}
\end{equation}

\textbf{Update $\mathbf{B}^{(i)}$.} $\mathbf{B}^{(i)}$ is divided into $\mathbf{\overline{B}}$ and $\mathbf{\widetilde{B}}_{i}$, and they are optimized separately. Remove all the irrelevant terms and update $\mathbf{\overline{B}}$, we have the following subproblem:
\begin{equation}
	\label{sub3}
	\begin{aligned}
		&\min _{\mathbf{\overline{B}}}
		\theta\left(\left\|\mathbf{\overline{B}}-\mathbf{V}\right\|_{F}^{2}\right)\\
		&s.t.\quad \mathbf{\overline{B}}\in\{-1,1\}^{r\times n_c}.
	\end{aligned}
\end{equation}
After some algebraic manipulation, this term can be transformed into a trace operation. It is equivalent to 
\begin{equation}
	\begin{aligned}
		&\max _{\mathbf{\overline{B}}} \operatorname{Tr}\left(\mathbf{V}\mathbf{\overline{B}}^{T}\right)\\
		&\text { s.t. } \mathbf{\overline{B}} \in\{-1,1\}^{r \times n_c}
	\end{aligned}
\end{equation}
Apparently, $\mathbf{\overline{B}}$ should be updated by 
\begin{equation}
	\mathbf{\overline{B}} = sign(\mathbf{V}).
	\label{Bc}
\end{equation}
Similarly, the optimal solution of $\mathbf{\widetilde{B}}_{i}$ is obtained by
\begin{equation}
	\mathbf{\widetilde{B}}_{i} = sign(\ddot{\mathbf{V}}\mathbf{R}^{(i)}).
	\label{B_tilde}
\end{equation}

By executing the preceding three stages, the objective function delineated by Eq. (\ref{overall}) is systematically minimized, culminating in convergence. For a given sample $\mathbf{m}$ from the $i$-th data modality, the corresponding hash code is determined through the computation $\mathbf{b} = \text{sign}(\mathbf{W}^{(i)} \cdot \phi(\mathbf{m}))$.

\section{Experiment}
\subsection{Datasets}
Two widely adopted cross-modal retrieval datasets are utilized for model evaluation: 1) The \textbf{MIRFlickr} dataset \cite{FLI}, which encompasses a compilation of 25,000 correspondences between images and texts, all procured from the Flickr service. Every correspondence is linked to a collection of tags, originating from a pool of 24 unique categories. From this dataset, we have curated a subset of 20,015 pairs, ensuring that each includes a minimum of 20 descriptive textual tags. In our setup, 10\% of the dataset is designated for the query collection, while the remainder forms the retrieval repository. 2) The \textbf{NUS-WIDE} dataset \cite{NUS}, a vast repository that includes 269,648 images paired with their respective textual annotations. We have focused on the most frequently occurring ten labels, resulting in a selection of 186,577 labeled samples. The imagery is encoded through the utilization of feature vectors based on a 500-dimensional bag-of-words model, whereas the textual data is encoded with 1,000-dimensional vectors. In this case, a single percent of the data corpus is allocated for the inquiry subset. It is important to note that, for training purposes, we have randomly extracted 10,000 paired samples from the aforementioned retrieval databases, combining them into a single training set. Within our experimental framework, the paired samples are drawn from the training set, with variations in the sampling ratio.

\subsection{Experimental Settings}
To verify the effectiveness of RREH, we selected several models from recent years and evaluated them in a semi-paired setting. They are PDDH \cite{pddh}, SPDH \cite{SPDH}, RUCMH \cite{RUCMH}, UAPMH \cite{UAPMH}, DAEH \cite{DAEH}, DUMCH \cite{DUMCH}, BATCH \cite{BATCH} and DAH \cite{DAH}. Among them, BATCH and DAH are supervised methods, and the rest are unsupervised methods. Our paper focuses on unsupervised semi-matching environments. Therefore, for fair comparison, pseudo labels are generated using K-means of these supervised methods \cite{CALM}. Completely paired models are trained using only paired data. For RREH, we set $\beta = 1e-2$ and $\theta = 1e-5$. The dimensionality for the kernelization process of both the visual and textual data is established at 500 and 1000, respectively. Across all datasets, the quantity of reconstruction anchor points for each modality is consistently configured to 600. We undertake two fundamental cross-modal retrieval operations: the first involves initiating queries with images to retrieve relevant texts, denoted as (${I \rightarrow T}$); the second commences with textual queries to identify corresponding images, represented as (${T \rightarrow I}$). 

\begin{figure}[t]
\centering
\begin{minipage}[b]{1.0\textwidth}
    \centering  
	\subfigure[${I \rightarrow T}$ on MIRFlickr] { 
		\includegraphics[width=0.38\columnwidth]{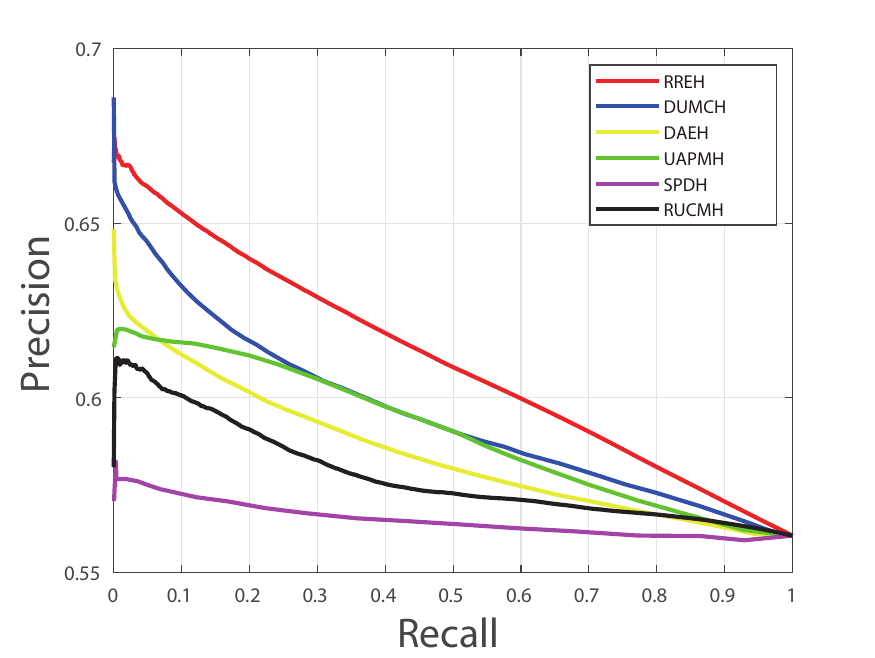}     
	} 
	\subfigure[${T \rightarrow I}$ on MIRFlickr] {
		\includegraphics[width=0.38\columnwidth]{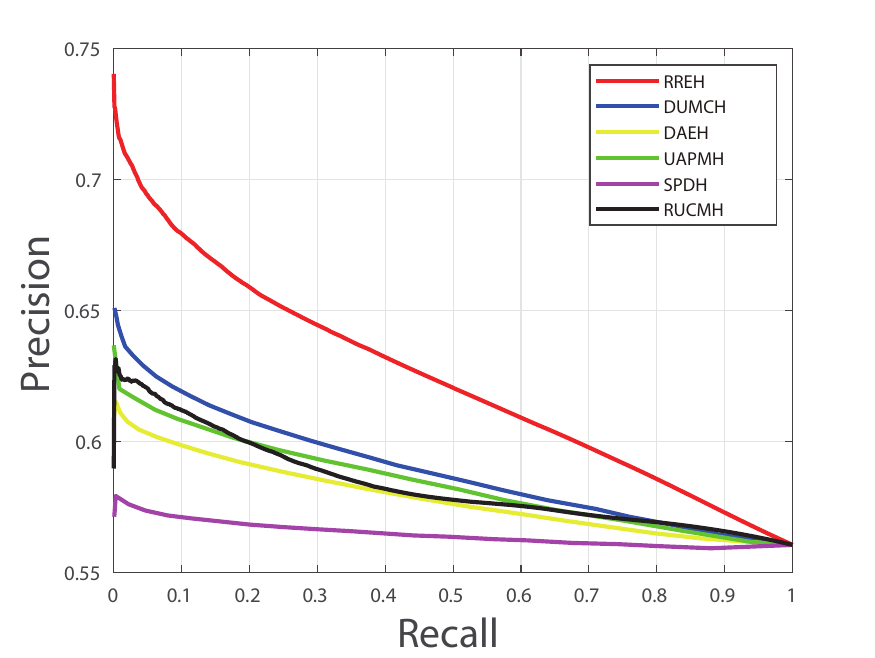}     
	} 
	\caption{The precision-recall curves of semi-paired models on MIRFlickr.}     
	\label{PR}         
\end{minipage}
\end{figure}


\begin{table*}[t]
	\caption{MAP results  on MIRFlickr (left) and NUS-WIDE (right) with 10\% samples paired. $*$ indicates that our results are statistically significant under the t-test (p $<$ 0.05) relative to the comparison model.}
    \renewcommand{\arraystretch}{1.1}
	\resizebox{\textwidth}{!}{
		\begin{tabular}{c  ccc  ccc  ccc  ccc}
			\hline
            Dataset&\multicolumn{6}{c}{MIRFlickr}&\multicolumn{6}{c}{NUS-WIDE}\\
            \hline
			\multirow{2}[1]{*}{Method}&\multicolumn{3}{c}{Images query Texts}&\multicolumn{3}{c}{Texts query Images}&\multicolumn{3}{c}{Images query Texts}&\multicolumn{3}{c}{Texts query Images}\\
			\cmidrule{2-4}\cmidrule{5-7}\cmidrule{8-10}\cmidrule{11-13}
			&16 bits&32 bits&64 bits&16 bits&32 bits&64 bits&16 bits&32 bits&64 bits&16 bits&32 bits&64 bits\\
			\hline
			PDDH \cite{pddh}&0.6468&0.6538&0.6592&0.6696&0.6798&0.6878&0.5511&0.5638&0.5696&0.5603&0.5654&0.5693\\
            BATCH \cite{BATCH}&0.5522&0.5648&0.5656&0.5652&0.5770&0.5717&0.4174&0.4168&0.4246&0.4309&0.4192&0.4276\\
			DAH \cite{DAH}&0.6008&0.5962&0.5990&0.5874&0.5933&0.5953&0.3846&0.3897&0.3901&0.3822&0.3841&0.3869\\
            \hline
			SPDH \cite{SPDH}&0.6194&0.5947&0.6105&0.6029&0.6003&0.6110&0.4838&0.4235&0.4570&0.4133&0.4091&0.4271\\
			RUCMH \cite{RUCMH}&0.6271&0.6199&0.6349&0.6309&0.6282&0.6291&0.4663&0.4794&0.4904&0.5117&0.5313&0.5274\\
			UAPMH \cite{UAPMH}&0.6203&0.6252&0.5925&0.6031&0.5990&0.5925&0.4973&0.4503&0.4649&0.4396&0.4364&0.4032\\
            DAEH \cite{DAEH}&0.6454&0.6040&0.6406&0.6183&0.6395&0.6716&0.4353&0.4805&0.4216&0.4399&0.4583&0.5213\\
            DUMCH \cite{DUMCH}&\textbf{0.6532}&\textbf{0.6569}&0.6602&0.6797&0.6853&0.6924&0.5382&0.5245&0.5268&0.5433&0.5476&0.5529\\ 
            RREH&0.6521&0.6554&\textbf{0.6608}&\textbf{0.6808}&\textbf{0.6914*}&\textbf{0.6965}&\textbf{0.5569*}&\textbf{0.5777*}&\textbf{0.5736*}&\textbf{0.5672*}&\textbf{0.5707*}&\textbf{0.5719*}\\
			\hline
		\end{tabular}
		\label{se_paired}}
\end{table*}
\vspace{-11pt}

\subsection{Performance Evaluation}
RREH is specifically designed for semi-paired data. To demonstrate the effectiveness of RREH on semi-paired data, we evaluate it using ten baselines. Among them, UAPMH was originally designed for multi-view image retrieval, and we adapted it for cross-modal retrieval. Noting that methods such as PDDH, BATCH and DAH adopt full-pairing settings in the original article. In our experiments, they also adopt semi-pairing settings and only use paired samples for training.

We conducted experiments using 10\% paired data and presented the MAP results for the above baselines in Table \ref{se_paired}. RREH generally outperforms all other methods, particularly in the Texts query Images task. RREH achieved comparable results to DUMCH on MIRFlickr and outperformed DUMCH on both tasks of NUS-WIDE. While UAPMH and RUCMH attained relatively good results on MIRFlickr, their performance significantly declined on NUS-WIDE. A crucial aspect is that the NUS-WIDE dataset significantly surpasses MIRFlickr in terms of volume, exhibits greater within-class diversity, and poses a challenge to obtain distinctive latent features merely through the alignment of corresponding instances, which also demonstrates the robustness of reconstructed relational embeddings. In addition, it can be seen from the results of RREH and PDDH that the semi-paired model we designed can also achieve comparable results to the fully paired supervised model. Among the fully paired methods, BATCH and DAH have the worst performance in the semi-paired setting. One possible reason is that the pseudo-labels generated by clustering may be inaccurate.

Furthermore, to evaluate the performance across varying proportions of correlated data, we have established the code length at 32 bits and the percentage of paired data to 60\%, presenting PR curves comparison for five unsupervised semi-paired hashing models as shown in Fig. \ref{PR}. RREH demonstrates superior performance compared to other methods. 

\begin{figure*}[t]
	\centering  
	\subfigure[$\beta$ on MIRFlickr] { 
		\includegraphics[width=0.4\columnwidth]{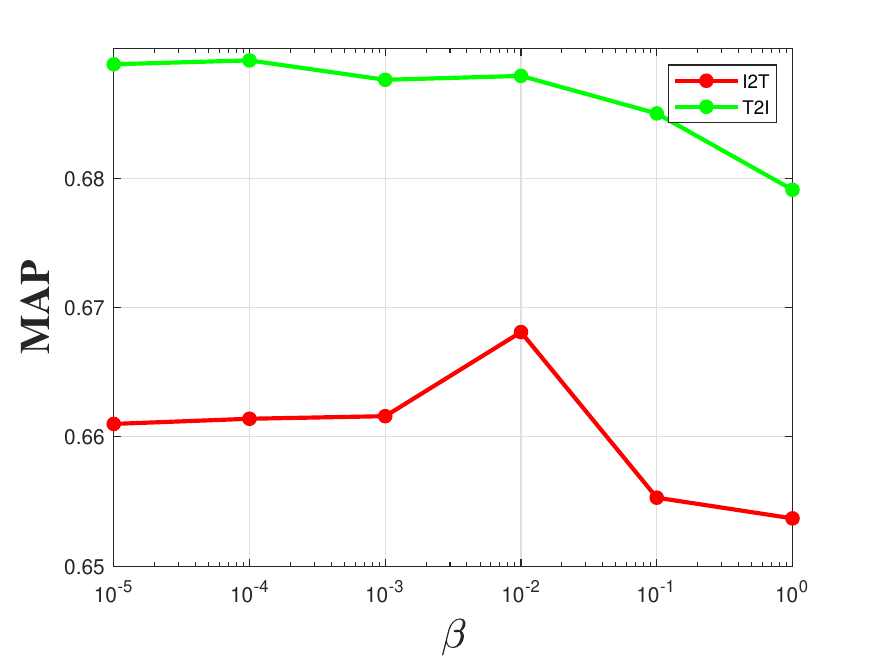}     
	} 
        \subfigure[$\theta$ on MIRFlickr] { 
		\includegraphics[width=0.4\columnwidth]{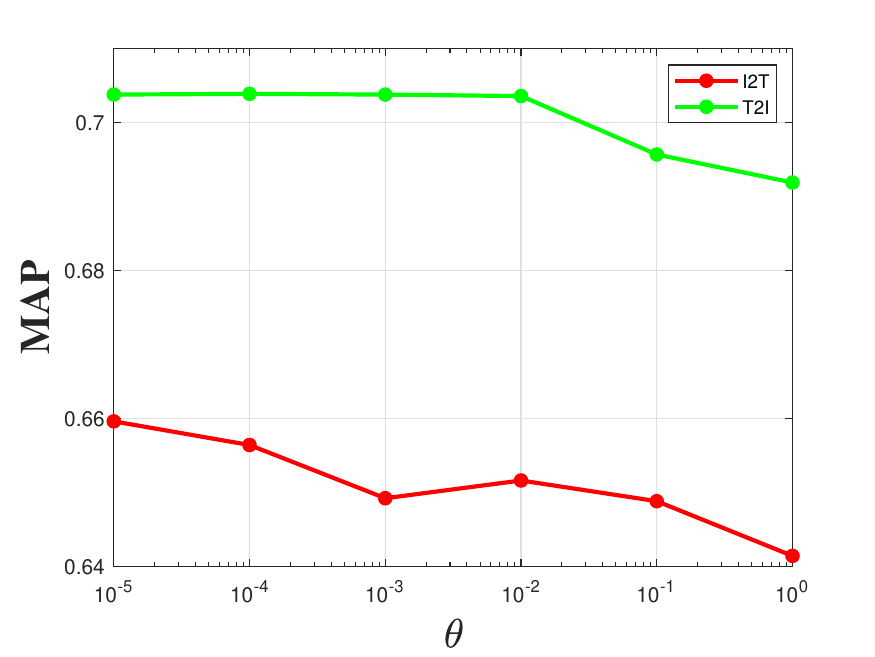}     
	}       
	\caption{Parameter analysis of $\beta$ and $\theta$ on MIRFlickr.}     
	\label{parameter}     
\end{figure*}

\vspace{-10pt}
\subsection{Parameter Analysis}
We also perform experiments to assess how sensitive RREH is to changes in hyperparameters. There are 2 hyperparameters in the model, i.e., $\beta$ and $\theta$. $\beta$ and $\theta$ both range from $1e-5$ to $1e0$. The length of the code is specified as 64. The outcomes are depicted in Fig. \ref{parameter}, illustrating the fluctuation of MAP in response to alterations in the two parameters. It is evident that the MAP values exhibit stability and a marginal enhancement, proving the effectiveness of the reconstruction relations embedding. The performance is alleviated with the increase of $\theta$ but drops drastically when $\theta$ is too large. One possible reason is that the hash codes are constrained to be binary and determined by the latent representation, while the latent representation is a continuous space, so large penalty on the hash code term compromises the learning of latent representation.

\begin{table*}[t]
	\caption{The MAP results of RREH and three variants with 5\% paired data.}
	\renewcommand{\arraystretch}{1.1}
	\resizebox{\textwidth}{!}{
		\begin{tabular}{c  ccc  ccc  ccc  ccc}
			\hline
            Dataset&\multicolumn{6}{c}{MIRFlickr}&\multicolumn{6}{c}{NUS-WIDE}\\
            \hline
			\multirow{2}[1]{*}{Method}&\multicolumn{3}{c}{Images query Texts}&\multicolumn{3}{c}{Texts query Images}&\multicolumn{3}{c}{Images query Texts}&\multicolumn{3}{c}{Texts query Images}\\
			\cmidrule{2-4}\cmidrule{5-7}\cmidrule{8-10}\cmidrule{11-13}
			&16 bits&32 bits&64 bits&16 bits&32 bits&64 bits&16 bits&32 bits&64 bits&16 bits&32 bits&64 bits\\
			\hline
   
			RREH-k&0.6343&0.6498&0.6448&0.6636&0.6737&0.6873&0.5536&\textbf{0.5661}&0.5639&0.5781&0.5687&0.5815\\
			RREH-r&0.6315&0.6367&0.6513&0.6359&0.6394&0.6537&0.5135&0.5129&0.5273&0.5311&0.5308&0.5373\\
            RREH-x&0.6246&0.6377&0.6406&0.6288&0.6396&0.6453&0.4889&0.4829&0.4868&0.5000&0.4939&0.4901\\
			RREH&\textbf{0.6498}&\textbf{0.6516}&\textbf{0.6585}&\textbf{0.6755}&\textbf{0.6866}&\textbf{0.6904}&\textbf{0.5594}&0.5606&\textbf{0.5696}&\textbf{0.5900}&\textbf{0.5770}&\textbf{0.5979}\\
			\hline
		\end{tabular}
		\label{ablation}}
\end{table*}
\vspace{-10pt}

\subsection{Ablation Study}
From the RREH model, three distinct variations have been developed, namely RREH-k, RREH-r, and RREH-x. The RREH-k variation forgoes the application of the kernel technique, opting instead to employ the raw data for model training. The RREH-r variant omits the reconstruction component, focusing solely on learning the hash function with the use of paired samples. Conversely, the RREH-k variant discards both the kernel method and the reconstruction learning process. An examination of Table \ref{ablation} reveals the Mean Average Precision (MAP) scores for RREH and its derivatives. It is evident that the RREH model outperforms its three variants across both retrieval tasks in terms of MAP. The RREH-k variant exhibits the poorest outcome, which underscores the significance of kernel techniques in capturing the intricacies of high-dimensional data. Furthermore, the superior performance of RREH over RREH-r validates the notion that reconstruction learning can mitigate the adverse effects associated with semi-paired data.
\vspace{-10pt}

\section{Conclusion}
In this paper, we proposed a novel unsupervised reconstruction relations embedded Hashing method, i.e., RREH. Instead of constructing a large-scale Laplacian matrix, RREH preserves data similarities with a linear reconstruction factor to generate discriminative hash codes. First, the reconstruction factors for all modalities are determined with original data. Then the reconstruction factors are embedded into latent representation to preserve data similarities. To obtain optimal hash codes and hash functions efficiently, an alternate optimization method is presented. Several experiments are conducted, proving the effectiveness of RREH  in terms of accuracy and efficiency. 
\vspace{-10pt}
\section{Acknowledgement}
This paper is supported by the Key Research and Development Program of Guangdong Province under grant No.2021B0101400003. Corresponding author is Xulong Zhang (zhangxulong@ieee.org) from Ping An Technology (Shenzhen) Co., Ltd..

%
%
%
\bibliographystyle{splncs04}
\bibliography{z_ref}
\end{document}